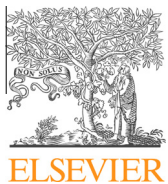

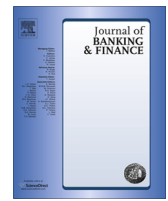

# Funding advantage and market discipline in the Canadian banking sector

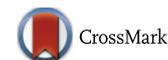

Mehdi Beyhaghi [a], Chris D'Souza [b], Gordon S. Roberts [c],*

[a] College of Business, University of Texas at San Antonio, One UTSA Circle, San Antonio, TX 78249-0631, USA
[b] Bank of Canada, 234 Wellington Street, Ottawa, ON K1A 0G9, Canada
[c] Schulich School of Business, York University, 4700 Keele Street, Toronto, ON M3J 1P3, Canada

## ARTICLE INFO

Article history:
Available online 29 August 2013

## ABSTRACT

We employ a comprehensive data set and a variety of methods to provide evidence on the magnitude of large banks' funding advantage in Canada in addition to the extent to which market discipline exists across different securities issued by the Canadian banks. The banking sector in Canada provides a unique setting in which to examine market discipline along with the prospects of proposed reforms because Canada has no history of government bailouts, and an implicit government guarantee has been in effect consistently since the 1920s. We find that large banks have a funding advantage over small banks after controlling for bank-specific and market risk factors. Large banks on average pay 80 basis points and 70 basis points less, respectively, on their deposits and subordinated debt. Working with hand-collected market data on debt issues by large banks, we also find that market discipline exists for subordinated debt and not for senior debt.

© 2013 Elsevier B.V. All rights reserved.

## 1. Introduction

Do banks face market discipline when they raise funds from wholesale deposits and bonds? This is an important question because current reform proposals aim to increase the incentive of bondholders to monitor banks more effectively instead of relying on costly government intervention to limit excessive risk taking by large banks. Market-oriented proposals to this end include mandatory subordinated debt, bail-ins and non-viability contingent capital (NVCC) (Evanoff et al., 2011; Evanoff and Wall, 2002; Basel Committee on Banking Supervision, 2010a, 2010b). Under the latter two proposals the debt-holders of a systemically important bank[1] face an administratively imposed or contractual partial conversion of debt into equity should the bank experience distress. NVCC forces the conversion of a bank's subordinated debt, while a bail-in extends NVCC and further enhances a bank's capital buffer by forcing the conversion of part of the banks' senior unsecured debt as well.

In this paper, we study the extent to which market discipline already exists in the Canadian banking sector. We also investigate if large Canadian banks have a funding advantage over other domestic banks after controlling for relevant risk factors, and finally we discuss whether and how a bail-in process would be practical in Canada.

The banking sector in Canada provides a unique setting in which to examine market discipline along with the prospects of enforcing a bail-in mechanism and NVCC because Canada has no history of government bailouts and an implicit government guarantee has been in effect consistently since the 1920s (Brean et al., 2011). In contrast, in the US Flannery and Sorescu (1996) and Balasubramnian and Cyree (2011) argue that market discipline is observed with error as implicit, too-big-to-fail (TBTF) guarantees have waxed and waned over time. In recent years, the perceived guarantee was undermined by the failure to bail out Lehman and again reinforced by subsequent rescues. This inconsistency in the US government approach to assisting large distressed banks makes Canada a more appropriate environment for study of the too-big-to-fail phenomenon.

In contrast with the US, the Canadian government has treated large banks consistently over an extended period of time. By examining anecdotal evidence, Brean et al. (2011) show that since the 1920s, the Canadian system has enjoyed stability grounded on an implicit guarantee for large banks. After the failure of a major bank in 1923, successive governments backed forced mergers as an alternative to failure of a weak bank. This was coupled with a TBTF implicit guarantee for surviving banks. During the Great Depression of the 1930s when deposit insurance did not exist in Canada, no banks failed despite deep market-value insolvencies (Kryzanowski and Roberts, 1993, 1999). As a result of this policy, combined with unrestricted national branching, the Canadian banking sector is dominated by the so-called Big Six banks.[2] At the end of our sample period at the fourth quarter-end of 2010, 28 domestic banks were active in Canada and the Big Six accounted for 93% of total assets. In recent years, banks in Canada have attracted favorable attention since they performed dramatically better than their peers in the United States. Canada did not experience a single bank failure during the financial crisis of 2007–2009.

---

[1] A systematically important financial institution (SIFI) is a financial institution whose collapse would pose a serious threat to the economy.

The present study employs a comprehensive data set and a variety of methods to provide evidence on the magnitude of the large banks' funding advantage in Canada in addition to the extent to which market discipline exists across different securities issued by the Canadian banks. To address the first objective, we measure the effective interest rate paid on different types of debt as well as credit spreads at the time of bond issuance. An effective interest rate for a debt item is calculated as the interest expense on that item, from a bank's income statement, divided by the level of that debt item from a bank's balance sheets in the same period. We control for market and bank-specific risk factors including various measures for equity (leverage), liquidity, and performance. In addition, we introduce controls for additional factors drawn from recent research.

Deposit withdrawals together with required interest rate increases are the two main vehicles available to debt-holders to prevent banks' excessive risk taking. Accordingly, we analyze the marginal impact of risk factors in order to determine whether interest rates act as a monitoring device. If so, changes in firm risk factors should immediately be reflected in a changing interest rate curve. Further, we examine the impact of bank risk on the growth of non-core, wholesale deposits. When market discipline is in force, we expect to see non-core deposit-holders withdraw their funds or (at a minimum) deposit less when the deposit-taking banks take on more risk, controlling for other factors.

In the next stage, we investigate the impact of seniority of a debt instrument on its sensitivity to issuers' risk factors for the "Big Six" systemically important banks.[3] For this purpose, we extract market data on debt (bond) issues. We expect junior debt to be more sensitive to the issuer's risk factors than more senior debt because in case of an asset liquidation, junior debt is paid only after senior debt. Pooling subordinated and senior debt in the same sample when examining sensitivity to bank risk (market discipline) could create a bias toward accepting the null hypothesis. If market discipline truly exists, we expect, first, the credit spread on each debt to be sensitive to the issuer's riskiness; and second, to the seniority of debt to be priced. If the market believes that some banks are so systemically important that the government would do anything to protect them from failure (TBTF), then the cost of raising debt for these banks must be generally lower and less sensitive to their riskiness.

We investigate the determinants of market credit spreads in various securities issued by the Big Six banks over a period of 21 years from 1990 through 2010 hand-collecting a comprehensive sample of their debt issuances and controlling for a host of variables, some of which are related to the banks' risk characteristics, while others reflect general market and economic conditions. We group issues by banks into senior and junior buckets using a five-stage bucketing algorithm after studying their contractual features including, but not limited to, their collateral types, maturity, stated seniority, coupon type, redemption features, and ratings. Our findings suggest that large banks enjoy a funding advantage over small banks after controlling for bank-specific and market risk factors. Large banks on average pay 80 basis points and 70 basis points less, respectively, on their deposits and subordinated debt. In general, bank debt in Canada is exposed to some degree of market discipline. Tests on levels and changes in the cost of debt, and also on wholesale deposit growth, show that during the sample period, the market reacts weakly to banks' risk taking and large banks have an advantage in terms of the effective interest rate they pay on debt and also enjoy more rapid non-core deposit growth.

In addition, the recent financial crisis provides an opportunity to examine if government actions dealing with failed/close-to-failure banks had an impact on banks' cost of debt. As discussed above, unlike the US, the Canadian government never rescued banks during our sample period (1990–2010). Rather, we examine the impact of a posited change in market perception of an implicit government guarantee on the risk-sensitivity of banks' cost of debt. Such a change in perception, we believe, might have occurred during the recent financial crisis, as there was likely less ambiguity in the market perception that the Canadian government would step in, as other members of the G7 did, and action become necessary to stabilize the market.[4] Our results show that bank-specific risk factors lose their significance in explaining funding costs during the crisis, especially for deposits. This finding is consistent with the argument that market awareness of government guarantees heightened during the crisis. We cannot, however, rule out an alternative explanation that during times of financial crisis returns on different assets tend to become positively correlated, because they all show together. Therefore, there might be less sensitivity to firm-specific risk factors across asset returns.

Working with hand-collected market data on debt issues by large banks, we also find that market discipline exists for subordinated and not for senior debt. This is important in light of the heavy reliance of banks on senior debt as a source of funding in the context of resolution plans that are based on the conversion of debt into equity or debt haircuts. The lack of sensitivity to banks' risk of large banks' senior-unsecured debt suggests that a bail-in might be appropriate in encouraging senior debt-holders to engage in monitoring banks more effectively. If senior debt-holders face credible losses once the bank is judged to be non-viable, there will be increased incentives to monitor bank risk taking ex ante and to charge riskier banks higher interest rates.

Our contributions to the literature are fourfold. First, we consider risk sensitivity (market discipline) for different levels of debt seniority. Second, this study encompasses most of the previous research in this field by drawing on both financial statements and market data to conduct its tests. Third, we shed light on the unique characteristics of Canadian banking, a system that is considered one of the soundest in the world. Fourth, our paper provides important policy implications for the design of bail-ins. The rest of this paper is organized as follows: the next section presents a brief review of relevant prior research followed by our methodology in Section 3 and data description in Section 4. In Section 5, we discuss our empirical results and Section 6 concludes.

---

[2] These are the Royal Bank of Canada, Canadian Imperial Bank of Commerce, Bank of Nova Scotia, Bank of Montreal, Toronto-Dominion Bank, and National Bank of Canada.

[3] We assume the Big 6 banks are the most systematically important banks in Canada due to their relative sizes. Among the Big 6, National Bank of Canada is the smallest. However, its total assets (equal to 153 billion Canadian dollars at the end of 2010) are more than the sum of assets of all other than big-6 domestic banks cumulatively. After the close of our sample period, Canada's federal banking regulator, the Office of the Superintendent of Financial Institutions, validated our assumption by naming the Big 6 as "domestic systemically important banks" and subject to a 1% surcharge on risk-weighted capital (Robertson, 2013).

[4] As will be discussed later, members of G7 stated they would take all necessary actions to stabilize financial markets after their October 2008 meeting.

## 2. Previous literature

Market discipline in bank-issued debt has been the focus of many studies over the past two decades (Flannery and Sorescu, 1996; Demirgüç-Kunt and Huizinga, 2004; Driessen, 2005; Krishnan et al., 2005; Caldwell, 2005, among others). Market discipline initiates when the probability increases that debt-holders of a bank will incur losses as banks take higher risks. As a result, debt-holders take action to penalize riskier banks by requiring higher rates of return or withdrawing their funds causing banks to act more prudently in order to avoid high costs of raising capital. The effectiveness of such market discipline critically depends on the ability of debt-holders to price changes in bank risk. Accordingly, research has investigated the extent to which banks' costs of raising debt reflect their riskiness (Avery et al., 1988). A part of this literature also highlights the impact of legislation and regulatory interventions on market discipline.[5] Flannery and Sorescu (1996) show that historically when the US government policy strengthens market perception that it would protect liability holders from credit losses, credit spreads on bank bonds show less sensitivity to bank-specific risk factors. In other words, investors have reflected changes in government policy toward absorbing private losses in the event of bank failure. Balasubramnian and Cyree (2011) demonstrate how government intervention in rescuing Long Term Capital Management (LTCM) in 1998 led to a TBTF perception by the market. They show that banks' cost of debt became less sensitive to bank-specific risk after this event.

Measurement of market discipline follows two streams of literature: the first uses effective interest rates derived from financial statements as a metric (Martinez-Peria and Schmukler, 2001; Demirgüç-Kunt and Huizinga, 2004, 2010; Hadad et al., 2010) while the second employs bond-issue credit spreads as a gauge of the cost of debt (Flannery and Sorescu, 1996; Krishnan et al., 2005; Balasubramnian and Cyree, 2011). In the first stream, bank characteristics, including their riskiness, cost structure and funding strategies, as well as general market conditions, explain the cost of debt. The second stream uses banks' risk factors, market factors, and issue characteristics as explanatory variables. Previous studies choose proxies for liquidity, equity capital, and performance as measures of riskiness. Our work seeks to integrate these two streams, we choose similar risk factors for both sets of tests drawn from factors identified in prior studies as important determinants of the cost of bank debt.

Previous studies employ various risk factors as controls including different measures for equity (leverage), liquidity, performance, and size. Balasubramnian and Cyree (2011) find that stock market volatility as a proxy for idiosyncratic risk adds explanatory power to their models. Demirgüç-Kunt and Huizinga (2010) also document relationships between the funding strategy or business model that banks employ and their risk with greater reliance on non-interest income and non-deposit funding, as observed in large and fast-growing banks, associated with increased bank risk. We control for these risk factors when we empirically examine banks in Canada. In addition to measuring the impact of risk factors on the cost of debt, we also examine the impact of a change in risk factors on changes in the cost of debt. If market discipline exists, we expect to see that variations in the cost of debt are explained by variations in risk factors. Krishnan et al. (2005) find significant results for the levels, but not the first differences. Balasubramnian and Cyree (2011) show that determinants of yield spread changes are jointly significant only for the period before

banks started issuing trust-preferred securities (TPS) in the Unites States.[6]

In summary, we draw on the previous literature in two ways: first, to identify the most relevant risk factors (bank-specific/market) that theoretically affect cost of debt for banks and, second, to calibrate banks' cost of debt.

## 3. Methodology

As explained above, to test for a funding advantage and market discipline for large Canadian banks, we examine both the effective interest rate that is implied by banks' financial statements and the cost of debt extracted from the market data in the form of credit spreads at the time of issuance. Working with financial statements facilitates broad comparisons as these statements are available in a standard format for all domestic banks. In contrast, the number of bond issues by other domestic banks is insignificant when one compares them with the Big Six. Fig. 1, Panel A presents the total amount of bond issues from Bloomberg for the nine largest Canadian banks. We use these market data (on bond issuance) to run our second group of tests in which we focus on large banks to determine if subordinated debt is more costly than senior debt and also whether subordinated debt is more sensitive to bank riskiness.

### 3.1. Funding advantage (large versus small banks)—financial statement measures

Our tests address how each bank's cost of issuing debt is determined by its riskiness (measured by accounting numbers), controlling for market-wide factors and government interest rates. Our methodology for this part follows Martinez-Peria and Schmukler (2001). In choosing risk factors, we follow the recent study by Demirgüç-Kunt and Huizinga (2010), adding a BIGSIX dummy set equal to one when a bank is one of the large six banks and 0 otherwise. We begin with a single-equation formulation as follows:

$$
\begin{aligned}
\text{Effective Interest Rate}_{b,t} = {} & A + Bt + C \; (\textit{BIGSIX Dummy}) \\
& + D(\textit{Market Risk Factors})_{t-1} \\
& + E \; (\textit{Bank Specific Risk Factors})_{t-1} \\
& + error_{i,b,t}
\end{aligned}
\tag{1}
$$

where $b$ represents each bank and $t$ represents the calendar quarter of observation. Bank-specific risk factors include equity, defined as the ratio of equity capital to assets as a proxy for bank capitalization; liquidity, the ratio of liquid assets to total assets, and performance measured by return on investments.[7] An overhead variable is constructed as the ratio of non-interest expense-to-assets to represent a bank's cost structure. Non-deposit funding, the ratio of non-deposit capital to total assets, represents banks' funding strategy. The effective interest rate is the ratio of quarterly interest expense on a debt item to its end-of-quarter balance from consolidated income statements and balance sheets. Effective interest rates are calculated for different types of liabilities including total debt, total deposits, wholesale deposits, and subordinated debt. Market risk factors reflect general market conditions such as interest rates on Canadian government debt and the unemployment rate. In robustness tests, we replace these with quarterly dummies. Further, following Flannery and Sorescu (1996), Krishnan et al. (2005) and Balasubramnian and Cyree (2011), who found that the impact of risk variables changed over time in the United States, we run separate models for the crisis period and a pre-crisis period.

---

[5] For a good review of this literature, see Covitz et al. (2004).

[6] TPS are junior to subordinated debt and similar to debentures and preferreds. TPS are generally longer-term, have early redemption features, make quarterly fixed interest payments, and mature at face value.

[7] The change in effective interest rates can be expected to be slow moving and not overly responsive to market or firm risk factors for slowly growing banks since such changes are primarily determined by the fixed rates on debt already in place for each bank. Such inertia creates a downward bias in the significance of the coefficients for our tests.

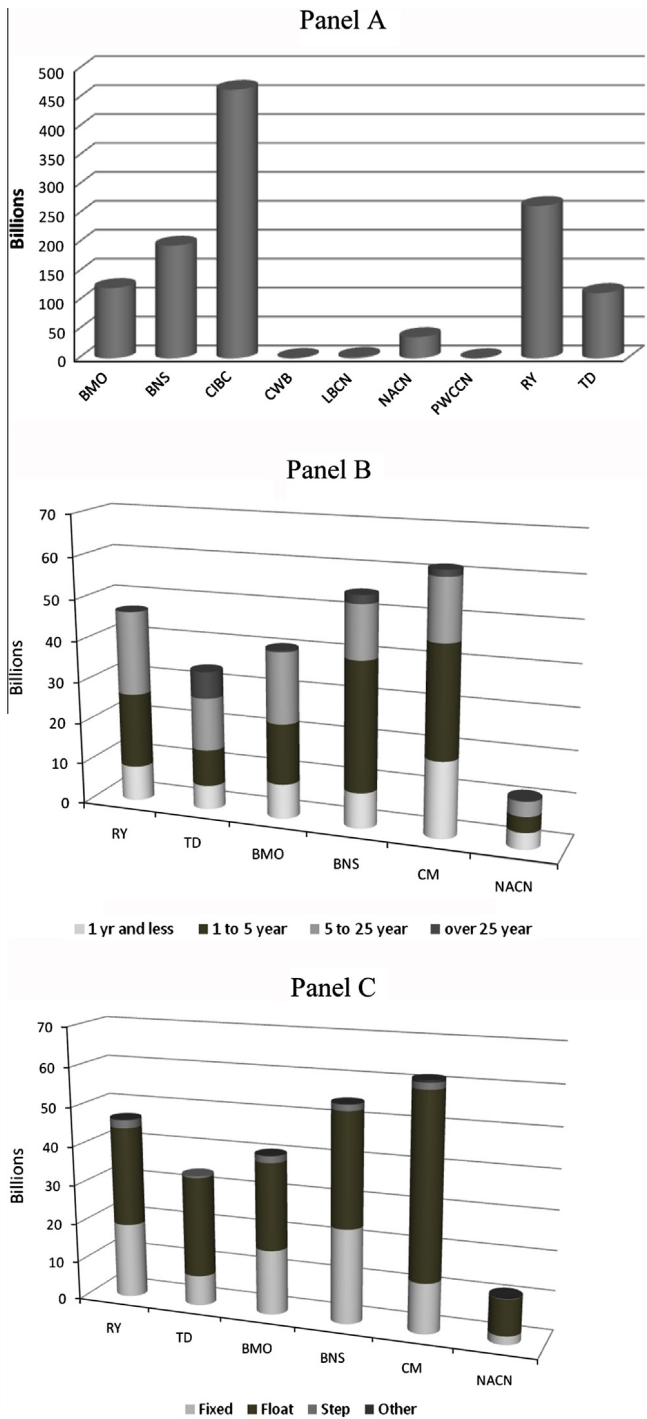

**Fig. 1.** Total amount of bond issuance by selected domestic Canadian banks during the sample period (1990–2010). Amounts are presented in billions of Canadian dollars. Panel A represents the total amount of bonds issued by the nine largest Canadian banks. Issuance amounts are converted into Canadian dollars when bonds are issued in different currencies. Issue amounts are then deflated/inflated to 2005 base prices. Panels B and C respectively show the year-to-maturity distribution and the coupon-type distribution of all CAD bond issuances by the so-called Big-Six Canadian banks during the period 1990–2010. Banks are presented with their tickers in the figure. These banks include Bank of Montreal (BMO), Bank of Nova Scotia (BNS), Canadian Imperial Bank of Commerce (CM), Canadian Western Bank (CWB), Laurentian Bank of Canada (LBCN), National Bank of Canada (NACN), Pacific & Western Bank of Canada (PWCN), Royal Bank of Canada (RY), and Toronto-Dominion Bank (TD). Caisse Centrale Desjardins, another large Canadian bank, is deleted due to its different legal structure. It is the representative of the Desjardins Group, which itself is an association of credit unions. The source of data is Bloomberg and the period covers 1990–2010.

We next employ a two-stage model to control for the impact of funding and maturity mix. Put another way, we want to rule out the possibility that differences in interest costs result from variety in funding mix and maturity rather than differences in bank riskiness. We first estimate expected interest costs on deposits as a function of their maturity and funding mix and use the residual as a spread variable in the second-stage analysis in which we are measuring the too-big-to-fail impact.[8] The first-stage is estimation of interest costs based on funding granularity as shown in Eq. (2) and the second stage is estimation of the interest cost residual, which is actual minus predicted interest cost, based on bank riskiness, BIGSIX status and other controls:

$$\textit{Effective Interest Rate}_{b,t} = A + B\,(\textit{Funding Mix}/\textit{Maturity Mix})_{t-1} + \textit{Residual Term}_{i,b,t}$$
(2)

$$\textit{Residual Term}_{i,b,t} = A' + B't + C\,(\textit{BIGSIX Dummy}) + D\,(\textit{Market Risk Factors})_{t-1} + E\,(\textit{Bank Specific Risk Factors})_{t-1} + error'_{i,b,t}$$
(3)

Eq. (3) is the same as Eq. (1) except that a spread variable is used instead of actual cost. Because funding and maturity mixes are available only for deposits, we limit our two-stage analysis to this interest cost category.

Returning to our original single-equation model, we conduct first-differences analysis to study the impact of *changes* in related risk factors on the *changes* in effective interest rates. Following Krishnan et al. (2005), we analyze the marginal impact of risk factors since one of our goals is to test whether interest rates act as a monitoring device. In this case, changes in firm risk factors should immediately be reflected in a changing interest rate curve. Our next models are as follows:

$$\textit{Change in Effective Interest Rate}_{b,t}$$
$$= A + Bt + C\,(\textit{BIGSIX Dummy})$$
$$+ D\,(\Delta\textit{Market Risk Factors})_{t-1}$$
$$+ E\,(\Delta\textit{Firm Specific Risk Factors})_{t-1} + error_{i,b,t}$$
(4)

where $\Delta$ represents change from $t - 1$ to $t$, such that:

$$\Delta\,\textit{Risk Factor}_t = (\textit{Risk Factor}_t / \textit{Risk Factor}_{t-1}) - 1$$
(5)

In addition, as a robustness check and following Martinez-Peria and Schmukler (2001), we employ quarterly growth in real deposits as an alternative measure of whether deposit-holders react to a bank's riskiness. In the presence of market discipline, we predict that an increase in the riskiness of a bank will negatively affect the growth in uninsured and/or non-core deposits. Uninsured deposits are not guaranteed by the Canada Deposit Insurance Corporation (CDIC), a federal Crown corporation created in 1967 and currently insuring deposits up to CAN 100,000. Core deposits are made by customers in the bank's general market area who tend to be loyal and consistent. Both core deposits and insured deposits are considered less sensitive to economic changes or bank riskiness than other bank debt. Detailed data for the portion of deposits that is uninsured or non-core are not available; however, we use wholesale deposits (deposits not open to individuals) as a proxy.[9] The model for this part has the following format:

---

[8] We thank the referee for suggesting this approach.

[9] The growth in deposits can be internal growth or growth by acquisition. In the latter case, a presumption might be created toward accepting the null hypothesis if a risky bank increased its wholesale deposits through acquiring another entity. Historical data, however, show that the overall size of the wholesale deposits that belong to acquired entities is negligible. This point notwithstanding, our empirical results for wholesale deposits are consistent with our other results.

*Real Growth in Wholesale Deposits$_{b,t}$*

$$= A + Bt + C \ (BIGSIX\ Dummy) + D \ (Market\ Risk\ Factors)_{t-1}$$
$$+ E \ (Firm\ Specific\ Risk\ Factors)_{t-1} + error_{i,b,t} \qquad (6)$$

The accounting data we use are at the consolidated level and include Canadian bank operations outside Canada, especially in the United States. Ideally, we should run the model only for domestic operations since banks differ in the ratio of foreign to domestic operations. In practice, however, the definition of foreign/domestic operations is not clear as domestic operations potentially include all transactions performed by domestic branches of Canadian banks, as well as all transactions in the domestic currency. Historical data show that both domestic and foreign branches of Canadian banks have issued debt in both domestic and foreign currencies and have sold it to foreign/domestic clients. Over the years, Canadian banks have issued debt in more than 20 currencies. While some distinctions among currencies are provided for, in reporting accounting items, the data include no breakdown between foreign and domestic branches. Due to integration of financial markets, there is no reason to believe that the market's perception of implicit government guarantees of large banks differs between foreign and domestic bondholders. Moreover, there is no indication in issued debt terms and conditions that Canadian and non-Canadian investors will be treated differently at the time of bankruptcy. Therefore, using data at the consolidated level should not cause any systematic bias when we examine funding advantages.

### 3.2. Market discipline for large banks—bond market yield spreads

Following Flannery and Sorescu (1996), Krishnan et al. (2005) and Balasubramnian and Cyree (2011), our variable of interest is each deal's issuance spread.[10] In contrast with these studies we consider only spreads at the time of issue because most of the securities in our sample are not actively traded in the secondary market. We use all available debt issues by banks to estimate the following regression for yield spread:

$$Spread_{i,s,b,t} = A + Bt + Cb + D \ (Market\ Risk\ Factors)_{t-1}$$
$$+ E \ (Firm\ Specific\ Risk\ Factors)_{t-1}$$
$$+ F \ (Issue\ Specific\ Characteristics)_{i,s,t} + error_{i,s,b,t} \qquad (7)$$

where $i$ represents each issue, $s$ represents different seniority level, $b$ is a bank fixed-effects dummy, and $t$ represents time of issuance. Bank-specific risk factors are similar to those in Eqs. (1) and (2) with the addition of bank asset size. We drop the Big Six dummy here as all our spread observations are from Big Six issuers. Market risk factors include those in Eq. (1) as well as the return on the market index (measured by the TSX index), market volatility (measured by the VIX index), and market liquidity measured by the slope of the term structure (10-year Treasury rates minus 2-year Treasury rates). In addition, issue-specific characteristics that affect yield spreads include redemption features such as callability, issue amount and bank size measured by log (Assets). We run this regression separately for different seniority levels and for different currencies in which banks have issued debt. We expect to see greater sensitivity to risk factors for less senior securities.

Krishnan et al. (2005) and Balasubramnian and Cyree (2011) conduct their tests on US bond transaction data from the National Association of Insurance Commissioners (NAIC) database from 1994 to 1999. These data allow the authors to construct a first differences model, in which changes in spreads are regressed on changes in risk factors. In the Canadian market, however, transaction data are not available for most banks in our sample because the secondary market is not active for most securities and there is no central database that reports bond transactions for those bonds that are traded. We employ only the market data on issue prices (explained in detail in the next section) hand-collected from Bloomberg. As a result, for each bond, we have one observation and, therefore, we cannot follow price changes in a specific security.[11]

## 4. Data and summary statistics

Historical bank financial statements, exchange rates, and Treasury rates for the Canadian economy are extracted from the database of the Bank of Canada. Market indicators including the VIX and TSX indices come from the Center for Research in Security Prices (CRSP). Detailed information related to debt issues is hand-collected from Bloomberg.[12] The advantage of Bloomberg over other sources is its comprehensiveness.[13]

Our sample includes all banks that have been active during recent decades but excludes trust and loan companies. We restrict the sample to the 1990–2010 period for two reasons: first, to be consistent with the available market data that we use for spread analysis and, second, since the required reporting format of financial statements has changed over time, using older data might cause inaccuracy in our analysis.[14] Further, we screen out subsidiaries of foreign and domestic banks since their funding strategies/costs are not independent of their parents. We also exclude all foreign banks and banks originally established under other jurisdictions. This leaves us with a sample of 2436 domestic bank-quarters. A total of 672 observations belong to banks that have become inactive during our sample period typically small banks that experienced difficulties and were then bought and merged into larger banks. Because such problem banks usually face a higher cost of funding before becoming inactive, including them in our sample could create a bias in favor of our hypothesis of a funding advantage for the Big Six.[15] Therefore, we omit all observations related to these banks. Our final sample includes 1764 observations representing six large and 15 small domestic banks.

In order to provide perspective on our financial statement data, Fig. 2 shows that deposits constitute the main source of funding for Canadian banks. On average, deposits are around 65% of large banks' and 83% of small banks' total liabilities and equity reflecting the tendency of smaller banks to follow a more traditional model of banking by funding a higher proportion of their assets through deposits. In addition, these figures show that studies that consider only bond spreads to examine whether large banks have a funding advantage may be incomplete as they ignore the largest sources of funding for banks. Fig. 2 also shows that, on average, smaller banks hold higher levels of equity than large banks. In addition, the average amount of subordinated debt across all groups of banks is at most 2%. This is important since regulators are considering Non-Viability Contingent Capital as a way of limiting larger banks' funding advantage and excessive risk taking. Under an NVCC, subordinated debt converts into equity at a trigger point of non-viability while the bank is still solvent, providing additional capital before taxpayers are involved. The current outstanding amount of subordinated debt (at most 2% of total assets) may not to be sufficient to recapitalize a failing bank in Canada.

---

[10] Other possible candidates that are used in industry or academia to measure the cost of raising debt are banks' ratings and CDS spreads. The former does not provide enough variation for testing while the latter is also not practical, as certain banks do not have an active CDS market. Therefore, we follow the most common methodology in the literature and use credit spreads as a proxy for cost of debt.

[11] Studies conducted in the US and that have used transaction data are limited to restricted samples and data sources. Most of these studies have used Warga, a dataset that has not been available since 1998. Balasubramnian and Cyree (2011) explain this in detail.

[12] Bank debt securities can be found within two of 10 market sectors in Bloomberg, namely Corporate Debt (CORP) and Preferred Shares (PFD). Other market sectors are government, mortgage-related, money market, municipal or state, equity, commodity, index, or currency securities. Bloomberg includes debt issued after January, 1981.

[13] Mergent and Thomson Reuter's SDC Platinum, for instance, do not provide data on short-term debt, i.e., maturity of one year or less. Mergent does not cover issues outside the United States. SDC also does not provide data on issues less than US$1 million. Other Canadian sources do not cover issuances by foreign subsidiaries of a Canadian bank.

[14] In 2009 there were around 30 domestic banks reporting to the Bank of Canada, there were also about 55 foreign banks active. Moreover, these exclude about 45 foreign and domestic banks that stopped reporting to the Bank of Canada during the period ending 2009. In addition there were around 60 institutions that reported to the Bank of Canada as Trust and Loan Companies.

[15] The banks that become inactive are typically riskier with lower performance ratios. Adding these banks to the sample would make our results even stronger, as they pay higher interest rates on their issues.

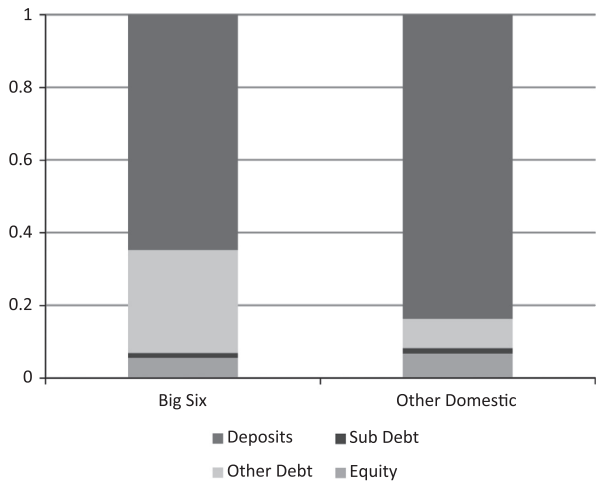

**Fig. 2.** Capital structure of Canadian banks based on consolidated balance sheets, quarter 4, 2010. This figure shows the average capital structure (liability and equity) of the Big 6 and other Canadian banks at the end of 2010. 'Other domestic' banks include 15 banks. Banks' sources of funding are dominated by deposits. Smaller banks demonstrate a more traditional model of banking and rely more on deposit funding than the Big 6. Subordinated debt represents at most 2% of banks' source of funding. Other debt includes cheques and other items in transit, advances from the bank of Canada, acceptances, non-controlling interest in subsidiaries, and other liabilities. Sample includes all active banks in Canada as of 2010, excluding subsidiaries and foreign banks, trust and loan companies and credit unions.

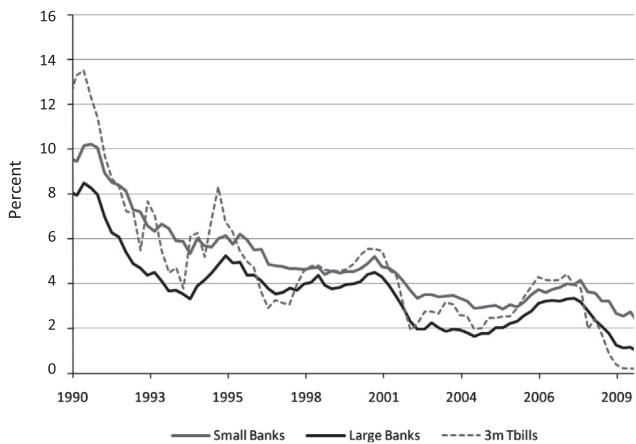

**Fig. 3.** Quarterly cost of debt for Canadian banks during the period 1990–2010. For each group of Canadian banks (Big 6 versus other domestic banks), the cost of debt is calculated as the median ratio of total interest expense during a quarter to the quarterly level of liabilities (also called implicit interest rate). Subsidiaries, foreign banks, loan and trust, and credit unions are excluded. This figure shows that Big 6 banks have a funding advantage in raising debt over other domestic banks (15 banks). Moreover, cost of debt is highly correlated with government rates (represented by quarterly (3-month) government of Canada treasury rates). *Data source*: Consolidated financial statements from Bank of Canada 1990–2010.

Fig. 3 represents the total cost of debt for Canadian banks over time for large and small banks and shows that, overall, the cost of debt has dropped from 1990 to 2010 consistent with the downward trend in interest rates. Also, Fig. 3 reveals that large banks on average have paid lower effective interest rates over time, supporting the hypothesis that they enjoy a funding advantage over smaller banks in the Canadian economy. We examine this finding in more detail in our multivariate analysis. Finally, the figure shows that the cost of debt is highly correlated with government rates, proxied by quarterly (3-month) treasury rates.

### 4.1. Market data

We hand collect detailed information for all debt issues by the 11 largest banks active in Canada by searching their tickers on Bloomberg.[16] Reported market data on the remaining domestic banks are negligible. This gives us a total of 12,224 issues that includes all debt issued (excluding preferred stock) by all banks' domestic and foreign branches on Bloomberg. We omit those issues that have missing or non-positive amounts leaving 10,267 observations. For reasons explained earlier, we restrict our sample to issues from the beginning of 1990 through the end of 2010 and this leaves us with a sample of 10,148 observations. A total of 339 observations belong to non-domestic banks and credit unions that report their financial statements under different regulations (the only institution with such characteristics is Caisse centrale Desjardins in our sample); 9805 observations are related to domestic banks.

Fig. 1, Panel A presents the total gross amount issued by each bank over the sample period across 25 different currencies. Issue amounts in other currencies are converted to Canadian dollars (CAN) using the exchange rate as of the issuance date and inflation-adjusted (base year 2005). Panel A shows that total bond issuance is dominated by the six large Canadian banks. The largest total gross amount other banks have issued is CAN 2352 million, by Laurentian, which is not comparable to the CAN 36,532 million issued by the smallest Big Six (National Bank of Canada). We focus exclusively on bond issues by the Big Six and this brings our sample size down from 9805 to 9766 issues.

The next step is to assign the issues to four buckets based on their seniority. If market discipline exists, one expects to see credit spreads sensitive to banks' specific risk factors. In other words, the market should react to the amount of risk a bank takes by requiring a higher return on investment when an issuer bank takes a higher risk and vice versa. Also, one expects to see junior claims exhibit greater sensitivity to an issuer's riskiness than senior claims, as their holders are more exposed to the risk of not receiving principal or coupon payments. Accordingly, we assign each debt issue to a bucket based on its seniority, which is its priority over other securities issued by the same issuer with respect to the payment of coupons and/or repayment of principal. We do this after studying the relevant contractual features reported on Bloomberg. Securities in different buckets differ in their potential for recovery in the event of an issuer's default.[17] We define four general buckets: se-cured, senior unsecured, subordinated, and junior subordinated. The last bucket has priority over preferred and common stocks.[18] The seniority bucketing procedure is fully explained in an appendix available on request from the authors.

After this procedure we end up with the following sample distribution along with each bucket's total amount (only the amounts related to the Big Six banks are reported): Bucket 1 (Secured): 31 deals, $CAN 38.4 billion; Bucket 2 (Senior Unsecured): 9224 deals, $CAN 1025.0 billion; Bucket 3 (Subordinated): 440 deals, $CAN 107.6 billion; and Bucket 4 (Junior Subordinated): 71 deals, $CAN 21.3 billion. As stated earlier, the relatively small amount of subordinated debt suggests that regulators would likely have to consider bail-ins as an additional tool to recapitalize Canadian banks.

### 4.1.1. Credit yield extraction

Credit yield spread is defined as the difference between a bond's yield to maturity and that of a corresponding Government of Canada security with the same time to maturity. Following Krishnan et al. (2005) and Balasubramnian and Cyree (2011), we use the cubic-spline interpolation method, which captures the non-linear aspects of the yield curve and extracts the entire daily yield curve of related Canadian securities. Related information for Canada rates comes from the central bank's publications.

In order to be consistent with our data source, we use Bloomberg's method of calculating yields to maturity at issue. Yield to maturity of a fixed-rate plain-vanilla (no redemption feature) coupon bond is calculated as the internal rate of return, considering market price at issuance, coupon rate, and time to maturity. Yields to maturity for floating/variable rate bonds are calculated assuming that future interest rates are based on a forward interest rate curve at the time of bond issuance and the bonds' expected cash flows. Spreads are then winsorized at 10% to eliminate any extreme values resulting from errors in reporting. A total of 2687 observations have all the required detail to calculate credit spreads.

### 4.1.2. Summary statistics on market data

To compare bonds in the senior and junior buckets and to measure their sensitivity to bank-specific and macro-risk factors, we focus exclusively on issues in Canadian dollars by the Big Six Canadian banks. We omit observations that belong to bucket 1 (senior secured) and bucket 4 (junior subordinated) for two reasons. First, the sample size in each bucket is not sufficient to perform statistical analysis. Second, the nature of securities in these buckets is different from those of buckets 2 and 3. The first bucket is secured, so spreads are influenced by the riskiness of the collateral and we do not have sufficient information to study the collateral. Securities in bucket 4 are also different in nature as most of them are perpetual. Therefore, comparing their yields to maturity to yields on bonds which mostly mature within 5 years could create inaccuracies in statistical analysis. Out of 2687 observations with valid spreads, 805 are issued in CAN and are either senior unsecured (bucket 2) or subordinated (bucket 3). Total issues across different maturity levels and coupon types are reported in Fig. 1, Panels B and C. Most of the issued debt matures within 5 years and has a fixed or a floating rate coupon.[19]

We also omit bonds with puttable or convertible redemption features. This brings our sample size down from 805 to 799. While we prefer to work only with plain-vanilla bonds with no redemption features, if we were to omit all callable securities, we would encounter a sample size problem. Therefore, we follow the related literature that has utilized the information conveyed by bonds with different redemption features (Bhojraj and Sengupta, 2003; Deng et al., 2007) and retain callable securities. We calculate their spreads, assuming the security is held to maturity, but control for the redemption features when we run the regressions.

**Table 1**
Summary statistics of all debt issues with calculated spreads by top 11 Canadian banks.

|  | Subordinated debt | Senior unsecured debt |
|---|---|---|
| Number of observations | 159 | 640 |
| *Credit spread (%)* |  |  |
| Mean | 0.35 | 0.19 |
| Median | 0.40 | 0.21 |
| Std. dev. | 0.53 | 0.57 |
| *Year to maturity* |  |  |
| Mean | 10.7 | 4.9 |
| Median | 10.0 | 5.0 |
| Std. dev. | 4.0 | 5.0 |
| *Issue amount ($m)* |  |  |
| Mean | 206.0 | 129.4 |
| Median | 100.0 | 20.0 |
| Std. dev. | 288.4 | 267.0 |

Table provides summary statistics on bond issuances that are used in spreads analysis regressions. The sample is restricted to those that are issued in CAD by the six largest Canadian banks from 1990 to 2010.

Table 1 provides summary statistics for all Canadian issues with calculated spreads used in our empirical analysis time to maturity, and issue amounts across different seniority levels. The results from Table 1 are consistent with the risk characteristics of the two main debt groups and verify the accuracy of our bucketing algorithm. Subordinated debt has higher mean and median spreads and a longer maturity. The number of subordinated issues is less than the number of senior unsecured issuances and average/median subordinated issuance carries a larger dollar value than average/median senior unsecured debt.

## 5. Empirical results

This section provides the results from the empirical analysis of the cost of debt derived from financial statements and market data. Table 2 presents dependent and explanatory variables that are used in the multivariate analysis.

### 5.1. Interest expense analysis based on financial statements

Table 3 shows single-equation estimates for interest paid on total debt, total deposits, wholesale deposits, and subordinated debt that are inferred from quarterly income statements and balance sheets (Eqs. (1) and (4)). Time effects are not reported to save space. Using these interest rates for cost-of-debt analysis has the advantage that they can be constructed for all reporting banks in the economy. A disadvantage, however, is that these interest rates might provide noisy estimates of actual costs of debt because each category (total debt, deposits, subordinated debt, etc.) is a combination of different securities with varying contractual features. In particular, deposits can be insured or uninsured and we know that insured deposits are likely less sensitive to risk factors. However, as explained above, there is no clear distinction between insured and uninsured deposits in Canadian bank financial statements. As a proxy for interest on uninsured deposits, we also use interest paid on wholesale deposits (deposits by non-individuals). This proxy is not completely accurate; while all deposits above CAN 100,000 are

---

[18] Junior subordinated debt is assumed to have the same priority as preferred shares in some classifications. They are mostly perpetual securities with high fixed coupon rates (similar to preferred stock), but since Bloomberg identifies preferred stocks under a different market sector, we treat these securities differently. All junior subordinated debt in our sample is callable.

[19] Less than 5% of bonds issued by Canadian banks in the sample have other coupon types such as zero-coupon, step coupon, etc.

**Table 2**
Explanatory variables used in the regressions (ratios are in decimals).

*Firm-specific variables*
– Equity (ratio of equity capital to total assets)
– Performance (return on assets)
– Liquidity (ratio of liquid assets to total assets)
– Overhead (the ratio of non-interest expenses to total assets)
– Non-deposit funding (the ratio of non-deposit capital to total assets)
– Log assets (natural logarithm of assets, inflation-adjusted)
– BIGSIX (a dummy variable that equals 1 if the issuer is one of the big 6 banks, 0 otherwise)
– Variance (variance of adjusted stock prices over the previous quarter)

*Market-specific variables*
– VIX index level
– Return on TSX index
– 3-Month treasury rates
– 10 year minus 2-year T-bill yields
– Unemployment rate

*Deposit characteristics (DN stands for demand and notice deposits. FT stands for fixed-term deposits)*
– DN-municipal and school corporations
– DN-deposit-taking institutions
– DN-individuals
– DN-other
– FT-federal and provincial
– FT-municipal and school corporations
– FT-deposit-taking institutions
– FT-individuals
– FT-other
– Less than 1 year
– Between 1 and 5 year
– More than 5 year

*Issue-specific characteristics*
– Coupon rate
– Log issue amount (inflation adjusted)
– Redemption feature (call)
– Coupon type (fixed, float)

not insured, not all deposits below this amount come from individual depositors. We refine this distinction in our two-stage tests below. Measurement error also exists in the subordinated debt category which is not broken down in financial reports but can include securities with different maturities and contractual features.

A further complication arises from the way our proxies for cost of debt are constructed from financial statements by dividing total interest paid on that category during a quarter (extracted from a bank's income statement) by the balance of that category of debt at the end of that period (extracted from the balance sheet). This calculation is affected by measurement errors, as a security might expire before a period ends and, therefore disappear from the end of period balance sheet, while its cost is calculated in the income statement. Similarly, a new security may be issued near a quarter-end, in which case the figure on the balance sheet would not be proportionate to the cost reported in the income statement. However, since this error is not systematic, we use implied interest expense analysis to infer whether large banks have a funding advantage here, and in the next section we focus on market data that are not affected by these problems.

### 5.1.1. Single-equation tests

The results in Table 3 show that different sources of funding respond to bank risk taking. In particular, the liquidity ratio has a significant negative effect on most costs of funding. The coefficient for return on investment is significantly negative for interest paid on subordinated debt, showing that profitable banks pay lower interest on their subordinated debt. As expected, government interest rates, measured by 3-month Treasury rates, directly and positively affect the cost of debt for banks. This is also supported by Fig. 3, which shows that the cost of debt in both small and big bank categories moves together with government rates.

The principal variable of interest in these models is the dummy variable BIGSIX and Table 3 shows that on average, Big Six Canadian banks pay around 80 basis points less for their deposits than other domestic banks. They also pay about 70 basis points less on their subordinated debt when one compares them with smaller banks. These findings support the hypothesis that big banks have a funding advantage over small banks.

The coefficients for two of the control variables, non-deposit funding and overhead, are of opposite signs for subordinated debt versus other types of debt (deposits), and significant. Greater dependence on non-deposit funding reduces the risk for deposit-holders while subordinated debt-holders require higher returns if banks rely more heavily on non-deposit sources of funding (1 basis point for a 1% increase in the ratio of non-deposit liabilities to total assets). The coefficient for performance, measured by return on investment, is significant only for subordinated debt with a negative sign as predicted. On average, everything else held constant, subordinated debt-holders charge a bank 20.6 bps less if the bank has shown 100 bps better performance (measured as net income to total assets) over other banks at the end of the previous period. The results suggest that subordinated debt-holders believe that better-performing banks are less likely to default on their debt and, therefore, are less risky.

**Table 3**
Large banks' funding advantage – single-equation regression results – levels of effective interest rates.

| | Interest on liabilities | | Interest on deposits | | Interest on wholesale deposits | | Interest on subordinated debt | | Real wholesale deposit growth | |
| --- | --- | --- | --- | --- | --- | --- | --- | --- | --- | --- |
| | Estimate | Std. dev. | Estimate | Std. dev. | Estimate | Std. dev. | Estimate | Std. dev. | Estimate | Std. dev. |
| Equity | −0.008 | 0.003 | 0.002 | 0.004 | −0.008 | 0.009 | −0.023 | 0.017 | 93.5 | 76.5 |
| ROA | −0.012 | 0.022 | 0.003 | 0.034 | 0.097 | 0.083 | −0.206** | 0.083 | −396.9 | 629.7 |
| Liquidity | −0.013*** | 0.002 | −0.012*** | 0.002 | −0.022*** | 0.004 | 0.043 | 0.007 | 71.5** | 34.8 |
| Overhead | −0.056** | 0.012 | −0.034** | 0.014 | 0.068* | 0.034 | 0.012*** | 0.071 | 69.6 | 267.5 |
| Non-deposit funding | −3.00E−04 | 0.002 | −0.007*** | 0.002 | −0.001 | 0.007 | 0.010** | 0.005 | −173.1*** | 59.6 |
| Unemployment rate | −0.002** | 7.00E−04 | −0.002*** | 0.001 | −0.003** | 0.001 | −0.001 | 0.002 | 27.8 | 11.9 |
| Big Six | −0.01 | 5.00E−04 | −0.008*** | 0.001 | −0.0006 | 0.001 | −0.007*** | 0.001 | 31.6** | 13.2 |
| 3 m T-bill rate | 0.002** | 3.00E−04 | 0.001*** | 3.00E−04 | 0.001** | 6.00E−04 | 0.001 | 0.001 | 2.6* | 5.6 |
| Intercept | 0.043*** | 0.006 | 0.049*** | 0.006 | 0.049*** | 0.012 | 0.051*** | 0.013 | −217.1* | 98.0 |
| Year fixed effect | Yes | | Yes | | Yes | | Yes | | Yes | |
| Adjusted R-square | 80.40 | | 82.88 | | 55.86 | | 50.23 | | 1.11 | |
| No of observations | 987 | | 947 | | 885 | | 804 | | 942 | |

Table presents the regression results on different proxies of the cost of debt and also a proxy for wholesale deposit growth across Canadian banks during the period of 1990–2010. Coefficient for year-fixed effects are not reported. Interest on different types of debt (total liabilities, wholesale deposits, total deposits, or subordinated debt) is calculated as the ratio of interest expense on that type of debt divided by the level of that type reflected in quarterly financial statements. Real wholesale deposit growth is a proxy for the growth in uninsured deposits. Definitions for the dependent and explanatory variables can be found in Table 2.
* Significant at 10%.
** Significant at 5%.
*** Significant at 1%.

**Table 4**
Large banks' funding advantage – single-equation robustness tests with quarterly dummies – levels of effective interest rates.

| | Interest on liabilities | | Interest on deposits | | Interest on wholesale deposits | | Interest on subordinated debt | | Real wholesale deposit growth | |
|---|---|---|---|---|---|---|---|---|---|---|
| | Estimate | Std. dev. | Estimate | Std. dev. | Estimate | Std. dev. | Estimate | Std. dev. | Estimate | Std. dev. |
| Equity | −0.007** | 0.003 | 0.002 | 0.004 | −0.009 | 0.010 | −0.022 | 0.018 | 87.4 | 78.1 |
| ROA | −0.011 | 0.022 | 0.010 | 0.034 | 0.099 | 0.086 | −0.209** | 0.088 | −374.2 | 651.4 |
| Liquidity | −0.013*** | 0.002 | −0.012*** | 0.002 | −0.022*** | 0.004 | 0.045*** | 0.008 | 72.8* | 35.5 |
| Overhead | −0.058*** | 0.012 | −0.035** | 0.014 | 0.068* | 0.036 | 0.006 | 0.074 | 76.7 | 274.5 |
| Non-deposit funding | −2.71E−04 | 0.002 | −0.007*** | 0.002 | 1.18E−04 | 0.007 | 0.011** | 0.005 | −173.9*** | 61.1 |
| Big Six | −0.010*** | 0.001 | −0.008*** | 0.001 | −0.001 | 0.002 | −0.007*** | 0.001 | 31.5** | 13.4 |
| Intercept | 0.030*** | 0.001 | 0.031*** | 0.001 | 0.024*** | 0.003 | 0.046*** | 0.004 | 7.8 | 23.2 |
| Quarter fixed effect | Yes | | Yes | | Yes | | Yes | | Yes | |
| Adjusted *R*-square | 80.53 | | 83.65 | | 54.20 | | 47.87 | | 1.73 | |
| No of observations | 987 | | 947 | | 885 | | 804 | | 942 | |

Table presents the regression results on different proxies of the cost of debt and also a proxy for wholesale deposit growth across Canadian banks during the period of 1990–2010. Quarterly dummies are used as substitutes for market risk factors and year fixed effects. The coefficients for Quarter fixed effects are not reported. Interest on different types of debt (total liabilities, wholesale deposits, total deposits, or subordinated debt) is calculated as the ratio of interest expense on that type of debt divided by the level of that type reflected in quarterly financial statements. Real wholesale deposit growth is a proxy for the growth in uninsured deposits. Definitions for the dependent and explanatory variables can be found in Table 2.
* Significant at 10%.
** Significant at 5%.
*** Significant at 1%.

The findings for overhead costs are counterintuitive. For subordinated debt and wholesale deposits, the coefficient is positive, meaning higher overhead cost increases the risk for these instruments. However, for total deposits and total debt, the coefficient is negative. The results are likely driven by core and insured deposits as higher overhead costs (e.g., personnel and building costs) decrease the margin for core/insured deposit-holders. Subordinated debt-holders and non-core deposit debt-holders, however, require higher returns, 1.2 and 6.8 bps respectively, per 1% increase in the ratio of overhead costs to total assets. Overall, the results for overhead costs support the notion that banks rely more heavily on core depositors to support overhead costs.

Our conclusion that Big 6 banks pay less for deposits might be subject to an alternative explanation: these banks draw retail deposits from their branch networks to a greater extent than smaller banks that rely principally on Internet deposits and deposit brokers. It would be desirable to find data on this reliance and to separate core deposits. Core deposits are considered a stable source of funds and are generally less sensitive to changes in short-term interest rates and bank risks than other forms of deposits. However, as discussed earlier, identifying core deposits is not practical considering the format in which Canadian banks report their deposits. Another way to address this is to use wholesale deposit growth as a robustness check. Table 3 also shows the results for real wholesale deposits' quarterly growth and reveals that banks with better liquidity, and banks that on average have higher overhead costs are able to attract different sources of funding and have higher wholesale deposit growth over time. Further, higher interest rates attract more deposits to the banks as holding cash becomes more costly for investors. Most importantly, demonstrating the robustness of our prior findings, Big Six banks attract more deposits over time and have a higher wholesale deposit growth.

To examine whether unobservable market risk factors affect the results in Table 3, we use quarter effects (dummy variables) to replace year effects and market risk factors in Table 4. The results in Table 4 show that Table 3's findings are generally robust to our choice of controls for market risks. In particular, Big Six has the same negative coefficient for interest on deposits and subordinated debt and positive sign for real wholesale deposit growth as previously. Further, there is now a negative sign on Big Six in the regression for interest on wholesale deposits which strengthens our results in Table 3.

*5.1.2. Two-stage tests*

Table 5 reports the results for funding and maturity mix analyses for deposits with the market characteristic controls employed in Table 4 (Eqs. (2) and (3)). While the proportions of insured and uninsured deposits are not available from financial reports, we have access to the mix of deposits in two categories: 1. Demand and notice deposits, and 2. Fixed-term deposits. Each of these categories is separately broken down further to (a) Federal and provincial, (b) Municipal and school corporations, (c) Deposit-taking institutions, (d) Individuals and (e) other. This enables us to investigate the impact of funding mix on deposit interest expense. In addition, the largest banks voluntarily report the maturity mix of their deposits when they discuss interest rate sensitivity in their annual reports.[20] Funding mix is available for all banks in our sample (Model 1 in Table 5, Panel 1). Maturity mix is available from annual reports for only 9 banks[21] from 1997 to 2010 (Model 2 in Table 5, Panel 1). Ideally we should use both funding mix and maturity mix variables in one regression, however, due to high correlations between these components inclusion of both sets of variables provides statistically unreliable results. The funding and maturity variables are in percentage formats.

The focus of this analysis is the coefficients in Panel 2. The results confirm our previous findings and show that our conclusions about funding advantage and market discipline are not driven by differences in funding/maturity mix. The coefficients in the two-stage analysis for the cost of deposits here bear the same signs and significance as those in our earlier single-equation regressions for a broader list of cost variables.

The only contradictory point is the coefficient on overhead costs in both models which is significantly positive for the funding mix regression (Model 1) and significantly negative for the maturity mix regression (Model (2)). As explained earlier, overhead proxies for bank's cost structure. Because the first model uses all the sample banks which is dominated by non-Big Six banks and the second model uses data only for 9 banks dominated by the Big Six, different signs might result from including banks with different cost structures in each sample.

**Table 5**
Two-stage tests of the impact of funding and maturity mix on interest on deposits.

| | Interest on deposits (1) | | Interest on deposits (2) | |
|---|---|---|---|---|
| | Estimate | Std. dev. | Estimate | Std. dev. |
| *Panel 1 – First stage* | | | | |
| DN-municipal and school corporations | −0.001[***] | 1E−4 | | |
| DN-deposit-taking institutions | −0.001[**] | 1E−4 | | |
| DN-individuals | −0.001[*] | 1E−4 | | |
| DN-other | −0.001[***] | 1E−4 | | |
| FT-federal and provincial | 0.003[***] | 3E−4 | | |
| FT-municipal and school corporations | 0.009[***] | 9E−4 | | |
| FT-deposit-taking institutions | −0.001[*] | 1E−4 | | |
| FT-individuals | −0.001[**] | 1E−4 | | |
| FT-other | −0.001[*] | 1E−4 | | |
| Less than 1 year | | | 2E−4[***] | 3E−5 |
| Between 1 and 5 year | | | 7E−5[*] | 4E−5 |
| More than 5 year | | | −0.005[*] | 5E−4 |
| Intercept | 0.129[***] | 0.042 | 0.025[***] | 0.003 |
| *R*-square | 22.31 | | 23.60 | |
| No of observations | 952 | | 448 | |

| | Residual of interest on deposits (1) | | Residual of interest on deposits (2) | |
|---|---|---|---|---|
| | Estimate | Std. dev. | Estimate | Std. dev. |
| *Panel 2 – Second stage* | | | | |
| Equity | −0.014[*] | 0.004 | −0.218[***] | 0.033 |
| ROA | 0.052 | 0.038 | −0.139 | 0.204 |
| Liquidity | 0.002 | 0.002 | −0.023[***] | 0.004 |
| Overhead | 0.088[***] | 0.016 | −0.407[**] | 0.205 |
| Non-deposit funding | −0.010[***] | 0.002 | −0.008 | 0.006 |
| Unemployment rate | −0.025[***] | 8E−4 | 9E−5 | 0.001 |
| Big Six | −0.003[***] | 6E−4 | −0.002[**] | 0.001 |
| 3 m T-bill rate | 0.001[***] | 4E−4 | 0.004[***] | 6E−4 |
| Intercept | −0.013[*] | 0.007 | 0.012 | 0.009 |
| Year fixed effect | Yes | | Yes | |
| Adjusted *R*-square | 72.85 | | 77.18 | |
| No of observations | 946 | | 447 | |

Table presents two-stage analysis of cost of deposits. In the first stage (Panel 1) the cost of deposit for each quarter-bank is regressed on percentage of each deposit component at the beginning of the period. Model (1) and Model (2) respectively estimate interest on deposit in terms of funding mix and maturity mix. Deposits are reported in two main categories: Demand and notice deposits, denoted by DT and Fixed-term deposits denoted by FT in the table. Each of these categories have 5 sub-categories. Subcategory DT-Federal and provincial is dropped to prevent multicollinearity of including independent variables, the sum of which is 100%. In Model (2) deposits are categorized into 4 groups based on their maturity: less than 1 year, between 1 and 5 year, more than 5 year and also other. The latter group includes interest insensitive deposits or floating rate and is not included as a dependent variable to prevent multicollinearity. All independent variables in Panel 1 are in percentage format. Panel 2 reports the results of the second stage regression, where the dependent variable is the residual from Panel 1. The funding mix is available for the sample period (1990–2010), however the maturity mix is only available from 1997 to 2010 for 9 banks.
[*] Significant at 10%.
[**] Significant at 5%.
[***] Significant at 1%.

To test the robustness of our two-stage model we repeat the test from our single-equation analysis replacing the year effects and market risk factors with quarterly dummies. The results (not reported) confirm that our findings are robust to this change.

In summary, these two-stage tests for deposits establish the robustness of the findings of our single-equation models. We next return to the single-equation setting for further robustness testing.

### 5.1.3. Impact of the financial crisis

During the financial crisis of 2007–2009, most countries with developed financial markets, the so-called G7,[22] provided explicit guarantees and/or bailout funding to the debt-holders of some of their large banks in order to stabilize financial markets. Moreover, in October, 2008, finance ministers pledged to take "all necessary steps" to help stem the crisis.[23] Although the Canadian government neither bailed out nor expressed any intention of bailing out a domestic bank, the market perception that the government would step in if necessary was likely heightened. Therefore, one might expect the cost of debt to become less sensitive to risk variables during the crisis. To examine this, we run separate regressions for the sub-period of the financial crisis (2007–2009) and for a preceding period of equal duration. The results in Table 6 strongly suggest that before the crisis, effective interest rates on deposits were significantly dependent on bank risk factors; however, during the crisis, this dependency largely disappeared. Liquidity, overhead, and non-deposit funding ratios impact the cost of raising deposits before, but not during the crisis. The results for subordinated debt are weaker, due to the inaccuracies in their cost of debt calculations as explained before. Table 6 also demonstrates that before and after the crisis, the

**Table 6**
Large banks' funding advantage– before and during the financial crisis period.

| Single-equation tests | Interest on deposits | | | | Interest on subordinated debt | | | |
| --- | --- | --- | --- | --- | --- | --- | --- | --- |
| | Pre-crisis (2004–2006) | | Crisis (2007–2009) | | Pre-crisis (2004–2006) | | Crisis (2007–2009) | |
| | Estimate | Std. dev. | Estimate | Std. dev. | Estimate | Std. dev. | Estimate | Std. dev. |
| Equity | 0.007 | 0.259 | −0.001 | 0.015 | −0.043 | 0.036 | −0.067 | 0.053 |
| ROA | −0.232 | 0.005 | −0.001 | 0.096 | −0.497** | 0.155 | −0.078 | 0.231 |
| Liquidity | −0.018*** | 0.005 | −0.006 | 0.005 | 0.031* | 0.018 | 0.025* | 0.014 |
| Overhead | −0.112*** | 0.031 | −0.070 | 0.073 | −0.028 | 0.159 | −0.236 | 0.179 |
| Non-deposit funding | −0.012** | 0.005 | −0.005 | 0.449 | 0.024* | 0.014 | 0.003 | 0.009 |
| Unemployment rate | −0.002 | 0.002 | −0.002 | 0.002 | −0.007 | 0.008 | 2.00E−3 | 0.003 |
| Big Six | −0.007*** | 0.001 | −0.009*** | 0.001 | −0.014*** | 0.003 | −0.001 | 0.002 |
| 3 m T-bill rate | 0.004*** | 0.001 | 0.207*** | 0.011 | 0.002 | 0.004 | 0.001 | 0.001 |
| Intercept | 0.038** | 0.018 | 0.054*** | 0.014 | 0.096* | 0.056 | 0.043* | 0.026 |
| Year fixed effect | Yes | | Yes | | Yes | | Yes | |
| Adjusted R-square | 71.69 | | 55.42 | | 23.23 | | 4.69 | |
| No of observations | 163 | | 147 | | 157 | | 157 | |

Table presents the regression results for effective interest rates on deposits and subordinated debt across Canadian banks over two equally long sub-periods. Coefficients for year-fixed effects are not reported. Interest on total deposits and subordinated debt is calculated as the ratio of interest expense for each of these types of debt divided by the level of that type reflected in quarterly financial statements. The definitions for the dependent and explanatory variables can be found in Table 2.
* Significant at 10%.
** Significant at 5%.
*** Significant at 1%.

**Table 7**
Large banks' funding advantage – single-equation regression results – changes in effective interest rates.

| | Change in interest on liabilities | | Change in interest on deposits | | Change in interest on wholesale deposits | | Change in interest on subordinated debt | |
| --- | --- | --- | --- | --- | --- | --- | --- | --- |
| | Estimate | Std. dev. | Estimate | Std. dev. | Estimate | Std. dev. | Estimate | Std. dev. |
| ΔEquity | −0.229*** | 0.020 | −0.129*** | 0.034 | −0.045 | 0.093 | −0.067 | 0.048 |
| ΔROA | −0.001 | 0.033 | −0.001 | 0.001 | −0.001 | 0.001 | 8.00E−5 | 0.001 |
| ΔLiquidity | −1.00E−04 | 2.00E−04 | −2.00E−4 | −2.00E−4 | 2.00E−5 | 6.00E−4 | 0.003 | 0.006 |
| ΔOverhead | 0.004* | 0.002 | −0.003 | 0.002 | 0.005 | 0.006 | −0.032 | 0.033 |
| ΔNon-deposit funding | −0.124*** | 0.033 | −0.171*** | 0.030 | −0.365*** | 0.089 | 0.060 | 0.040 |
| ΔUnemployment rate | −0.404* | 0.146 | −0.030 | 0.132 | 0.005 | 0.340 | −0.046 | 0.128 |
| Big Six | −0.017* | 0.009 | −0.013 | 0.008 | −0.061*** | 0.023 | −0.006 | 0.009 |
| Δ3 m T-bill rate | 0.075*** | 0.025 | 0.056*** | 0.023 | 0.207*** | 0.011 | 0.063*** | 0.004 |
| ΔIntercept | −0.048** | 0.020 | 0.049*** | 0.006 | −0.072* | 0.043 | −0.035* | 0.017 |
| Adjusted R-square | 15.41 | | 15.39 | | 44.16 | | 27.48 | |
| No of observations | 967 | | 926 | | 865 | | 784 | |

Table presents the first difference regression results on different proxies of the cost of debt across Canadian banks during the period of 1990–2010. Coefficients for year-fixed effects are not reported. Interest on different types of debt (total liabilities, wholesale deposits, total deposits, or subordinated debt) is calculated as the ratio of interest expense for that type of debt divided by the level of that type reflected in quarterly financial statements. A change in variable $X_t$ (also represented by $\Delta X_t$) is defined as $(X_t / X_{t-1}) - 1$. Definitions for the dependent and explanatory variables can be found in Table 2.
* Significant at 10%.
** Significant at 5%.
*** Significant at 1%.

Big Six banks paid lower interest rates, controlling for relevant risk variables. On another note, adjusted R-squared shrinks from the pre-crisis to the crisis sub-period across both deposits and subordinated debt categories, confirming the finding that firm and market risk factors play a less significant role in explaining variations in costs of raising these categories of debt during the crisis. In unreported regressions, results using quarterly dummies as a replacement for market risk factors and year dummies further support findings in Table 6.

As mentioned earlier, the result that effective interest rates are less sensitive to bank-specific risk factors during the crisis might have an alternative explanation. Anecdotal evidence shows that during a financial crisis, different asset prices tend to move together. Therefore, as the probability of a systemic breakdown increases, bank-specific risks might be less important in investors' perceptions. Nonetheless, since the change in coefficient significance is more observable for deposits than for subordinated debt, one can infer that there is less ambiguity among deposit-holders that their investment will be protected by the government than there is among subordinated debt-holders.

### 5.1.4. Changes in effective interest rates

Results for the first difference regressions for the single-equation model are in Table 7. Changes in interest rates are regressed on changes in risk factors across different quarters, following Eq. (4). The results show that risk factors are playing a marginal role in changes in interest rates. Interestingly, the coefficient for the change in equity becomes significant for total liabilities and deposits while, earlier, in Tables 4 and 5, the coefficients for equity were insignificant. A larger positive change in equity is associated with a greater decrease in the cost of debt, supporting the notion that higher levels of equity make a bank less risky. The significance of the change in equity, but not the level could arise from regulatory restrictions on the level of equity whereas periodic changes in equity might be due more to bank-specific situations. To analyze this further, in unreported regressions we introduce the change in equity as an additional variable in Tables 3 and 4. The sign and significance of the results do not change, and the coefficient for change in equity is insignificant. In the sole case of interest on total liabilities (Table 3, first model, when we add the change in equity, the coefficient for the level of equity becomes negative

**Table 8**
Market discipline in Big Six banks' bond issuance – regression results.

| | Subordinated debt | | Senior unsecured debt | |
|---|---|---|---|---|
| | Estimate | Std. dev. | Estimate | Std. dev. |
| *Bank characteristics* | | | | |
| Log(Assets) | −1.323[*] | 0.728 | −0.270 | 0.218 |
| Equity | −42.679[*] | 22.20 | −6.922 | 8.358 |
| ROA | 50.879 | 105.2 | 18.236 | 29.118 |
| Liquidity | −1.602 | 2.855 | −0.203 | 1.184 |
| Ratio of non-deposit funding | 0.611 | 2.750 | −0.792 | 1.008 |
| Overhead | 27.156 | 94.68 | −2.555 | 29.841 |
| Variance | −0.004 | 0.023 | 0.004 | 0.005 |
| *Bond characteristics/liquidity* | | | | |
| Fixed | 0.177 | 0.123 | 0.216[***] | 0.047 |
| Coupon rate | 0.119[**] | 0.058 | 0.055[***] | 0.013 |
| Log(Issue Amount) | −0.040 | 0.036 | 0.018 | 0.014 |
| Callable | 0.563[***] | 0.172 | 0.316 | 0.051 |
| Time to maturity (Years) | −0.006 | 0.012 | 0.010[*] | 0.006 |
| Coupon frequency (per Year) | 0.003 | 0.013 | −0.018[***] | 0.005 |
| *Market conditions* | | | | |
| 10 year minus 2 year Treasury Rates | −0.030 | 0.206 | −0.026 | 0.080 |
| 3-Month Treasury Rates | 0.013 | 0.103 | 0.146[***] | 0.047 |
| Unemployment Rate | 0.053 | 0.246 | −0.121 | 0.098 |
| Average VIX Index | 0.003 | 0.019 | 0.014[***] | 0.004 |
| Return on TSX Index | 0.768 | 1.097 | −0.461 | 0.325 |
| Year Fixed Effect | Yes | | Yes | |
| Bank Fixed Effect | Yes | | Yes | |
| Intercept | 28.584[**] | 14.212 | 6.745 | 4.473 |
| Number of Observations | 133 | | 554 | |
| Adjusted R Square | 24.84 | | 59.71 | |

In table credit spreads of all issues in Canadian dollars by Canadian banks are regressed on bank-specific risk factors, bond/liquidity specific characteristics and market-specific factors. Regressions are controlled for year- and issuer-fixed effects. For a list of dependent and explanatory variable definitions, see Table 2.
[*] Significant at 10%.
[**] Significant at 5%.
[***] Significant at 1%.

and significant (the significance and signs of other variables do not change). This suggests that, overall, the total cost of debt decreases as the level of equity becomes higher after controlling for shocks to the level of equity.

In addition, Table 7 shows that banks' funding strategies matter in explaining changes in effective interest rates. A positive change in the ratio of non-deposit funding is associated with a negative change in the cost of raising deposits and total liabilities, but has no significant impact on the cost of subordinated debt.

Most importantly, Table 7 supports the hypothesis that Big Six banks enjoy a funding advantage. The coefficients for the BIGSIX dummy are negative, which means that changes in spreads are, in general, smaller for the largest banks. That is, after controlling for risk factors, the costs of debt for large banks change less than those of the smaller banks. Also, changes in government interest rates, measured by 3-month T-bill rates, are the most important factor in explaining changes in effective interest rates, an observation that is supported by previous literature and theoretical models.

Comparing adjusted *R*-squared of different regressions, we find that, in general, levels are explained better than changes consistent with Krishnan et al. (2005). For instance, the adjusted *R*-squared for the level of interest on liabilities regression in Table 3 is 80.4%, whereas the same statistic for the first difference (change) regressions for interest on liabilities is 15.41% in Table 7. Among first difference regressions, changes in wholesale deposits and subordinated debt are better explained by changes in risk factors (44.16% and 27.48%, respectively) than changes in other types of debt.

In summary, the results from Tables 3–7 show that, in general, banks in Canada are exposed to a degree of market discipline and that the Big Six Canadian banks have a funding advantage over other domestic banks. The next section provides a different type of analysis by using market data on bond spreads.

### 5.2. Bond credit spreads analysis

In this section, we report results from estimation of a fixed-effects OLS model for all issues in Canadian dollars across two main seniority buckets. Credit spreads are regressed on market and bank-specific risk factors, and issue characteristics. If market discipline exists, we expect to observe significant sensitivity of yield spreads to bank-specific risk factors. In coordination with the empirical analysis we performed in the previous section, we use the common set of explanatory variables defined in Table 2 and discussed above.

Bank-specific risk variables are lagged one period and represent the most recent information that bond investors have with respect to banks' accounting statements. We control for size, performance (ROA), liquidity,[24] and equity capital calculated from financial statements. In addition, following Balasubramnian and Cyree (2011), we utilize the variance of stock prices (adjusted for splits by Bloomberg) over the prior period as a further control variable. The other risk factors are consistent with the first sets of tests that we run for effective interest rates. Additional variables (in Balasubramnian and Cyree) that we do not use are either highly correlated with the ones we include or unavailable for Canadian banks.

Spreads are also affected by general market conditions and banks may time the market by issuing debt when demand is high. We include five market factors to control for market conditions all measured at the most recent quarter end before issuance. Market volatility is measured as the average level of the Chicago Board Options Exchange Market Volatility Index (VIX) over 30 trading days before the security issuance date. Market return is the return on the TSX index (the main stock market index in Canada) over the 30 trading days before issuance date. Three-month Treasury rates represent the government interest rate, and 10-year minus 2-year Treasury rates is a measure for liquidity in the market inferred from the term structure of government rates.

---

[24] Liquid assets are measured as the sum of cash and cash equivalents, securities issued or guaranteed by Canada/Canadian provinces/Canadian municipal or school corporations and non-mortgage loans, less an allowance for impairment restricted to call and other short loans.

**Table 9**
Market discipline – large versus small banks – single equation tests.

| | Interest on liabilities Big six banks | | Interest on liabilities other banks | |
|---|---|---|---|---|
| | Estimate | Std. dev. | Estimate | Std. dev. |
| Equity | −0.114 | 0.067 | −0.026[**] | 0.009 |
| ROA | −0.372 | 0.249 | 0.039 | 0.046 |
| Liquidity | −0.015[*] | 0.007 | −0.011[*] | 0.005 |
| Overhead | −1.164 | 0.857 | −0.065[***] | 0.022 |
| Non-deposit funding | −0.018[**] | 0.005 | 0.004 | 0.005 |
| Log (Assets) | 0.001 | 5E−4 | −0.001[***] | 4E−4 |
| Intercept | 0.025[*] | 0.011 | 0.051[***] | 0.007 |
| Bank fixed effect | Yes | | Yes | |
| Quarter fixed effect | Yes | | Yes | |
| Adjusted R-square | 94.33 | | 78.02 | |
| No of observations | 480 | | 507 | |

Table presents the results of estimating interest on total liabilities separately for Big Six banks and other banks in Canada during the period of 1990–2010. The structure of regressions is similar to the ones in Table 5. However log (Assets) as a measure of bank size and bank fixed effects are also considered in this table. Definitions for the dependent and explanatory variables can be found in Table 2.
[*] Significant at 10%.
[**] Significant at 5%.
[***] Significant at 1%.

Bond-specific features follow Balasubramnian and Cyree (2011), Bhojraj and Sengupta (2003) and Deng et al. (2007) to control for bond riskiness and liquidity and include issue amount, coupon payment frequency, redemption features (callability), time to maturity and whether the bond is a fixed-rate or a floating-rate issue. These features can impact the demand for a bond and thus its cost of issue and liquidity. In addition, coupon rates control for tax effects.[25]

We report the regression results in Table 8 for senior unsecured and subordinated debt issues. As stated above, all the observations here are from Big Six banks and consequently we drop the Big Six dummy. The regressions include year and issuer fixed effects. The coefficients for two bank-specific risk factors (size and equity) are statistically significant for the subordinated issues but none of the bank-specific variables are significant for the senior unsecured issues. This suggests that Big Six Canadian banks are exposed to bond market discipline exclusively via their subordinated debt. Investors in the subordinated debt market adjust their required rate of return after observing the issuer's level of equity capital over the previous quarter and banks with higher equity capital have lower credit spreads. Further, Table 8 shows that even among Big Six banks, larger banks (as measured by the log of assets) pay lower interest rates on their subordinated debt. Our results also confirm that market conditions are important for senior issues. When markets are volatile, as measured by the level of the VIX index, the cost of debt goes up since lenders require higher rates of return. The results also show that fixed-rate bonds have higher spreads consistent with their greater exposure to interest rate risk relative to floating rate bonds. Finally, the coefficient for coupon rates is significantly positive as in Balasubramnian and Cyree (2011) reflecting the higher tax attracted by higher coupon bonds.[26]

In summary, the results regarding Big Six banks' subordinated debt in this section provide some support for the existence of market discipline; however, senior debt-holders do not price bank riskiness in their required rate of return. This supports the too-big-to-fail argument and the notion that investors believe in an implicit government guarantee at least in the senior unsecured debt sector.

### 5.3. Market discipline analysis

Given our finding that market discipline exists for the Big Six, it should be present for non-Big-Six banks to a greater degree. Because market data for bonds issued by non-Big-Six banks is scarce or not available, we return to our earlier setting in which we derive interest costs from financial statements. We now measure interest on total liabilities separately for the Big Six and others. The set up is similar to the one in Table 3, with the Big Six dummy excluded as the regressions are sorted by Big Six and others. In addition we use a measure of bank size and control for bank fixed effects.

The results in Table 9 show that both Big Six and other banks are exposed to market discipline as coefficients for selected measures of bank risk attain significance in both regressions suggesting that bank risk affects the cost of debt. The effect is more observable for other (non-Big Six) banks. Of interest is the coefficient for bank size, measured as log (Assets). Table 9 shows that the cost of debt is sensitive to bank size for smaller, non-Big Six, banks but not for the Big Six banks providing further support for our TBTF argument. An alternative explanation is that size variations may be greater for smaller banks over the sample period.

### 6. Conclusion

Larger banks are effectively shielded from market discipline if the market perceives that they will be bailed out in times of distress. To alleviate moral hazard problems and to control banks' excessive risk taking, proposed reforms seek to minimize government intervention and to enhance the incentive of bondholders to monitor banks more effectively. Two major proposals designed specifically for the resolution of failed/close-to-failure large banks with emphasis on market mechanisms are non-viability contingent capital (NVCC) and debt bail-in. Under the former, subordinated debt converts into equity (upon a trigger at the point of non-viability) providing additional capital before taxpayers become involved. A bail-in extends NVCC to enhance a bank's capital buffer, forcing senior creditors to bear losses by contributing to the recapitalization and, hence, to the resolution of a failed (or weak) institution. Given its critical role in these proposed reforms, it is important to measure the extent to which market discipline already exists.

---

[25] Balasubramnian and Cyree also use a dummy for sinking-fund bonds. The number of such bonds in our data after the screening process described in the methodology section above is insignificant.
[26] In the previous section, we performed a pre-versus post-crisis analysis. It would be interesting if we could conduct the same type of analysis for the spreads; however, our sample size restricts us from doing so. In our screened sample, there are only 18 subordinated debt issues during the crisis period (2007–2009) and 54 for the pre-crisis period.

We study the extent of market discipline in the Canadian banking sector and also whether large banks enjoy a funding advantage over smaller banks. Canada provides a unique setting in which to examine market discipline as there have been no government bailouts in the history of the banking sector yet an implicit government guarantee has been in effect since the 1920s (Brean et al., 2011). Canada is also an interesting case, as Canadian banks performed dramatically better than their peers in the United States, despite the integration of the two economies. Our results show that in Canada, large banks enjoy a funding advantage and on average pay 80 and 70 basis points less for deposits and subordinated debt, respectively, controlling for various risk factors. Taken together, these findings support the notion that the market believes in an implicit government guarantee in the Canadian banking sector. They also suggest that market discipline exists weakly in Canada. Debt-holders respond to bank-specific risk factors, measured by ratios of equity capital, liquidity, performance, cost structure, and business models by adjusting the interest rate they require from banks or by withdrawing their funds. The results are robust to the introduction of added controls for Big Six banks funding and maturity mixes. Findings from examining credit spreads in large bank bond issues suggest that, unlike for subordinated bonds, the credit spreads for senior bonds are not significantly sensitive to bank-specific risk factors.

Further, it is important to recognize how different contractual features of securities can affect the value of contingent capital (Pennacchi (2011), Sundaresan and Wang (2010), McDonald (2010) and Glasserman and Nouri (2010), among others.) Senior debt includes securities with different coupon types, maturities, and redemption features and, therefore, it is critical to identify clearly which securities are appropriate to be considered under a bail-in mechanism and can provide additional capital at the time of distress. Regulators would like to exclude securities with maturities of less than 1 year in order to limit runs by money-market investors. Since Canadian banks' outstanding debt typically matures within 1–5 years, excluding shorter maturity debt (up to 1 year) would not significantly alter the availability of securities subject to bail-in according to statistics provided in Fig. 1 Panels B and C. Also, given the relatively high outstanding amounts of fixed and floating rate instruments, regulators would have to consider the bail-in treatment for each group. Forcing similar haircuts on floating and fixed-rate instruments could have different implications for each unintentionally creating preferential treatment to one class of creditors within a seniority bucket.

More broadly, these results support the argument that a bail-in mechanism can potentially enhance market discipline by engaging senior debt-holders more effectively in the monitoring of large banks. Further, we also find that market discipline weakened during the financial crisis as the market perception that the government would step in if necessary was likely heightened. However, we cannot reject an alternative explanation that during the crisis asset prices moved together and, therefore, their sensitivity to firm-specific risk variables dropped.

In addition, our results have implications for how regulators could implement a bail in. In particular, the current amount of subordinated debt for Canadian banks (at most 2% as reflected on balance sheets) is unlikely to generate sufficient capital at the time of distress. Therefore, considering senior debt, as suggested under a bail-in mechanism, seems to provide a more effective resolution outcome.

## Acknowledgements

We are grateful to the referee for this journal for suggestions that greatly improved the paper. We also received helpful comments from conference participants at the 2012 International Banking, Economics and Finance Association/Western Economic Association International Meetings in San Francisco, the 2012 Northern Finance Association Meetings in Niagara Falls, the 2012 Australasian Finance & Banking Conference in Sydney, the 2013 Southwestern Finance Association Meetings in Albuquerque, and seminar participants at the Bank of Canada, Concordia University and York University. We also thank Deniz Anginer, James Chapman, Evren Damar, Douglas Evanoff, Xing Huang, Lawrence Kryzanowski, Alexandra Lai, Nadia Massoud, Adi Mordel, Carol Ann Northcott and Claudio Wewel for helpful comments. Rucha Deshmukh, Sheisha Kulkarni, Amit Soni, and Jie Zhu provided outstanding research assistance. Mehdi Beyhaghi thanks the Bank of Canada for support during his visit. Gordon Roberts acknowledges financial support from the Social Sciences and Humanities Research Council (SSHRC) of Canada. The views expressed in this paper are those of the authors. No responsibility for them should be attributed to the Bank of Canada.